\documentclass[twoside]{article}
\usepackage{bezier,amsfonts}

\def\noi{\noindent}

\makeatletter

\renewcommand{\thesubsubsection}%
        {\arabic{section}.\arabic{subsection}.\arabic{subsubsection}.}
\renewcommand{\@oddhead}{\raisebox{0pt}[\headheight][0pt]{%
   \vbox{\hbox to\textwidth{\rightmark \hfil \rm \thepage \strut}\hrule}}}
\renewcommand{\@evenhead}{\raisebox{0pt}[\headheight][0pt]{%
   \vbox{\hbox to\textwidth{\thepage \hfil \leftmark \strut}\hrule}}}
\newcommand{\heads}[2]{\markboth{\protect\small\it #1}{\protect\small\it #2}}
\newcommand{\Acknow}[1]{\subsection*{Acknowledgement} #1}
\makeatother

\newcommand{\Title}[1]{\noi {\Large #1} \\}
\newcommand{\Author}[2]{\noi{\large\bf #1}\\[2ex]\noi{\it #2}\\}
\newcommand{\Abstract}[1]{\vskip 2mm \begin{center}
     \parbox{16.4cm}{\small\noi #1} \end{center}\bigskip}
\newcommand{\foom}[1]{\protect\footnotemark[#1]}

\newcommand{\email}[2]{\footnotetext[#1]{e-mail: #2}}
\def\sect{Sec.\,}

\def\nqq{\hspace{-2em}}
\def\nhq{\hspace{-0.5em}}

\def\cm{\hspace{1cm}}
\def\inch{\hspace{1in}}
\def\mas{\mbox{$\mathstrut$}}

\def\eq{Eq.\,}
\def\eqs{Eqs.\,}
\def\beq{\begin{equation}}
\def\eeq{\end{equation}}
\def\bear{\begin{eqnarray}}
\def\al{&\nhq}
\def\lal{&&\nqq {}}               
\def\bearr{\begin{eqnarray} \lal}
\def\ear{\end{eqnarray}}
\def\earn{\nonumber \end{eqnarray}}
\def\tst{\textstyle}
\def\dst{\displaystyle}

\def\nn{\nonumber\\ {}}

\def\nnn{\nonumber\\ \lal }

\def\yy{\\[5pt]}

\def\eql{\al =\al}

\def\eqdef{\stackrel{\rm def}{=}}
\def\e{{\,\rm e}}
\def\d{\partial}

\def\sign{\mathop{\rm sign}\nolimits}
\def\diag{\mathop{\rm diag}\nolimits}

\def\const{{\rm const}}

\def\half{{\tst\frac{1}{2}}}

\newcommand{\vars}[1]{\left\{\begin{array}{ll}#1\end{array}\right.}

\newcommand{\toas}[1]{\mathop {\longrightarrow}\limits_{#1}}

\def\ep{\epsilon}
\def\o{\omega}
\def\A{{\cal A}}
\def\M{{\cal M}}

\def\S{{\cal S}}
\def\V{{\cal V}}
\def\oI{{\overline I}}
\def\od{{\overline d}}
\def\oD{{\overline D}}

\def\olam{\overline{\lambda}}
\def\ophi{\overline{\varphi}}
\def\uc{{\underline c}}
\def\vY{\vec Y{}}
\def\vZ{\vec Z{}}
\def\hq{\hat q{}}
\def\Som{\S_{\o}}
\def\sumo{\sum_{\o}}
\def\summo{\sum_{\mu\in\Som}}
\def\sumn{\sum_{i=1}^{n}}
\def\sums{\sum_s}
\def\umx{u_{\max}}
\def\m{{\rm m}}
\def\Nsq{N_s^2}
\def\Noq{N_\o^2}
\def\Fei{F_{\e I}}
\def\Fmi{F_{\m I}}

\newcommand{\Theorem}[2]{\medskip\noi {\bf #1. \ }{\sl #2}\medskip}
\newcommand{\eps}{ \varepsilon }
\newcommand{\R}{\mathbb R}
\newcommand{\N}{\mathbb N}

\def\mO{{[-1]\mathstrut}}
\def\pO{{[1]\mathstrut}}

\def\Nsq{N_s^2}

\def\Ref#1{Ref.\,\cite{#1}}
\def\rank{\mathop{\rm rank}\nolimits}

\def\sph{spherically symmetric\ }

\def\bhs{black holes}
\def\wh{wormhole}
\def\whs{wormholes}

\heads{K.A. Bronnikov}
{Gravitating Brane Systems: Some General Theorems}

\begin{document}
\thispagestyle{empty}
\begin{center}
               RUSSIAN GRAVITATIONAL SOCIETY\\
               INSTITUTE OF METROLOGICAL SERVICE \\
               CENTER OF GRAVITATION AND FUNDAMENTAL METROLOGY\\
\end{center}
\vskip 3ex
\begin{flushright}
                                         RGS-VNIIMS-11/98 \\
                                         gr-qc/9806102
\end{flushright}

\vskip 2ex

\Title
{GRAVITATING BRANE SYSTEMS: \yy SOME GENERAL THEOREMS}
\medskip

\Author{K.A. Bronnikov\foom 1}
{Centre for Gravitation and Fundamental Metrology, VNIIMS,
3-1 M. Ulyanovoy St., Moscow 117313, Russia}

\Abstract
    {Multidimensional gravity interacting with intersecting electric and
	magnetic $p$-branes is considered for fields depending on a single
	variable.  Some general features of the system behaviour are
	revealed without solving the field equations. Thus, essential
	asymptotic properties of isotropic cosmologies are
	indicated for different signs of spatial curvature; a no-hair-type
	theorem and a single-time theorem for \bhs\
	are proved (the latter makes sense in models with multiple time
	coordinates). The validity of the general observations is verified for
	a class of exact solutions known for the cases when certain vectors,
	built from the input parameters of the model, are either orthogonal in
	minisuperspace, or form mutually orthogonal subsystems. From the
	non-existence of Lorentzian \whs, a universal restriction is obtained,
	applicable to orthogonal or block-orthogonal subsystems of any
	$p$-brane system.  }

\email 1 {kb@goga.mainet.msk.su}

\section{Introduction}   

	In the weak field limits of the bosonic sectors of supergravities
	\cite{SS}, superstring
     and M-theory, their generalizations and modifications [2--6]
     there naturally appear
     multiple self-gravitating scalar dilatonic fields and antisymmetric
     forms, associated with $p$-branes.

     This paper continues the studies of
     such models on the basis of a general action, see (\ref{2.1}), without
	fixing the total space-time dimension $D$ or other input parameters
\cite{IMO}--\cite{bobs},
     thus to a large extent abstracting from the details of specific
	underlying models, but with a hope to predict some
	features of new models, unformulated by now.  We will here deal with the
	one-variable sector of the model, where all fields depend
	on a single coordinate: time in cosmological models, a radial
	coordinate in \sph models, etc. In this case the model reduces to a
	Toda-like dynamical system in minisuperspace, see (\ref{3.12}),
	(\ref{3.16}).

     Much work has been devoted to searches for exact solutions and their
     subsequent analysis.
     Thus, in Ref.\,\cite{bim97}, the most general one-variable solution was
     presented for the case when certain vectors $Y_s$ in the target space,
     built from the input parameters of the model, form
     an orthogonal system (OS). This solution describes a set of
     intersecting electrically and magnetically charged $p$-branes and
     generalized many previously found ones, beginning with
     Schwarzschild and Reissner-Nordstr\"om and ending with dilatonic and
     some more special $p$-brane solutions
(\cite{bshik,LPTX,br95,AIR,LPX}, etc.).
     The OS solution was further generalized \cite{bobs} to models where
     $Y_s$ form a block-orthogonal system (BOS). The OS solution is
	recovered when each block consists of a single vector.
	Other families of exact solutions have been found for cases
     when $Y_s$ form bases of integrable Toda models, see
	\cite{IM6,LPX,LuMaha} and references therein. Many solutions are known
	beyond the one-variable sector (\cite{IMmapa} and references therein).

	The exact solutions have disclosed
	many features of interest of physically relevant configurations, such as
	cosmological models and \bhs. The generality of these features
	remains, however, questionable, since the equations of motion can be
	solved exactly only for special (though numerous) choices of the input
	parameters. To have an idea of what can and what cannot be
	expected from yet unknown solutions, it makes sense to try to extract
	some information directly from the equations. Such an attempt is
	undertaken here.

     It appears possible to reveal some important properties of $p$-brane
     cosmologies, namely, the nature of asymptotics for
	different signs of spatial curvature. For \sph configurations, among
	other results, two theorems about \bhs\ (BHs) are proved:  a ``no-hair
	theorem", that a BH is incompatible with the so-called quasiscalar
	$F$-forms (see (\ref{2.25})), and a ``single-time theorem", that even
	in spaces with multiple times a black hole may only exist with its
	unique, one-dimensional (physical) time.

     One more general observation is \cite{bobs} the absence of \sph
     Lorentzian \whs\ under the requirement that all the fields bear
     positive energy, just as in conventional general relativity.
	On the other hand, for the known families
     (OS and BOS \cite{bobs}) of exact solutions one can deduce necessary
	and sufficient conditions under which a specific solution describes a
	\wh, no matter, Lorentzian or Euclidean.  Combined, these results lead
	to a universal restriction upon the input parameters of the model,
	valid for any brane system which has an OS or BOS subsystem (Theorems 4
	and 4a, already announced \cite{bobs} in a slightly different form).
	Having been obtained on the basis of specific exact solutions, this
	restriction still applies to systems for which solutions are yet to be
	found.

     The paper is organized as follows. The introductory \sect 2
	describes the model and a convenient
	Toda-like representation of its one-variable sector, in line with our
	previous papers. \sect 3 is devoted to general properties of
	cosmological and \sph $p$-brane configurations. \sect 4 gives a brief
	description of the OS and BOS solutions, necessary for
	obtaining the above-mentioned universal restriction. The latter is
	formulated in \sect 5. \sect 6 contains some concluding remarks, in
	particular, on the use of different conformal frames.

\section{The model. Minisuperspace representation} 

     The starting point is, as in Refs.\,[10--15], the model action for
	$D$-dimensional gravity with several scalar dilatonic fields
	$\varphi^a$ and antisymmetric $n_s$-forms $F_s$:
\beq                                                         \label{2.1}
     S = \frac{1}{2\kappa^{2}}
	             \int\limits_{\M} d^{D}z \sqrt{|g|} \biggl\{
	R[g]
	- \delta_{ab} g^{MN} \d_{M} \varphi^a \d_{N} \varphi^b
                                                    - \sum_{s\in \S}
 	\frac{\eta_s}{n_s!} \e^{2 \lambda_{sa} \varphi^a} F_s^2
                  \biggr\},
\eeq
     in a (pseudo-)Riemannian manifold
     $\M = \R_u \times \M_{0} \times \ldots \times \M_{n}$
	with the metric
\beq
     ds^2 = g_{MN}dz^M dz^N =
            w \e^{2{\alpha}(u)} du^2 +                    \label{2.11}
    	        \sum_{i=0}^{n} \e^{2\beta^i(u)} ds_i^2, \cm  w=\pm 1,
\eeq
     where $u$ is a selected coordinate ranging in $\R_u \subseteq \R$;
     $g^i = ds_i^2$ are metrics on $d_i$-dimensional
     factor spaces $\M_i$ of arbitrary signatures $\eps_i=\sign g^i$;
     $|g| = |\det g_{MN}|$ and similarly for subspaces;
     $F_s^2 =  F_{s,\ M_1 \ldots M_{n_s}} F_s^{M_1 \ldots M_{n_s}}$;
     $\lambda_{sa}$ are coupling constants;
     $\eta_s = \pm 1$ (to be specified later);
     $s \in \S$,  $a\in \A$, where $\S$ and $\A$ are some finite sets.
     All $\M_i$, $i>0$ are assumed to be Ricci-flat, while $\M_0$ is allowed
     to be a space of constant curvature $K_0= 0,\ \pm 1$.

	In the one-variable sector, $\varphi^a = \varphi^a(u)$.
	The set of indices $\S = \{s\}$ in (\ref{2.1})
	will be used to jointly describe essentially
     $u$-dependent electric ($\Fei$) and magnetic ($\Fmi$) $F$-forms, to be
     associated with different subsets $I = \{ i_1, \ldots, i_k \}$
     ($i_1 < \ldots < i_k$) of the set of numbers labelling the factor
     spaces:  $\{i\} = I_0 = \{0, \ldots, n \}$. Thus one can write
\beq                                                     \label{2.23}
     \S = \{s\} = \{\e I_s\} \cup \{\m I_s\}.
\eeq
    A given $F$-form may have several essentially (non-permutatively)
    different components, both electric and magnetic;  such a situation is
    sometimes called ``composite $p$-branes" \cite{AIR}%
\footnote{There is an exception:  two components
   having only one noncoinciding index, cannot coexist since in this case
   there emerge nonzero off-block-diagonal components
   of the energy-momentum tensor (EMT), while the
   Einstein tensor in the l.h.s. of the Einstein equations is
   block-diagonal.  See more details in \Ref {IM4}.}.
    For convenience, we will nevertheless treat essentially different
    components of the same $F$-form as individual (``elementary") $F$-forms.
    A subsequent reformulation to the composite ansatz is straightforward.

    So, by construction, nonzero components of $\Fei$ carry coordinate
    indices of $u$ and the subspaces $\M_i,\ i\in I$, those of $\Fmi$ ---
    the indices of $\M_i,\ i\in \oI \eqdef I_0 \setminus I$ simce a magnetic
    form is built as a form dual to a possible electric one. Therefore
\beq
    n_{\e I} = \rank F_{\e I} = d(I) + 1,\cm
    n_{\m I} = \rank F_{\m I} = D- \rank F_{\e I} = d(\oI)  \label{2.22}
\eeq
    where $d(I) = \sum_{i\in I} d_i$ are the dimensions of the subspaces
    $\M_I = \M_{i_1} \times \ldots \times \M_{i_k}$.

    Several electric and/or magnetic forms (with maybe different coupling
    constants $\lambda_{sa}$) can be associated with the same $I$ and are
    then labelled by different values of $s$.  (The index $s$ by $I$ is,
    however, sometimes omitted when this cannot cause confusion.)

    This problem setting covers various classes of models:
    isotropic and anisotro\-pic cosmologies, $u$
    is a timelike coordinate and $w=-1$;
    static models with various spatial symmetries (spherical, planar,
    pseudospherical, cylindrical, toroidal), where $u$ is a spatial
    coordinate, $w=+1$, and time is selected among $\M_i$; and
    Euclidean models with similar symmetries or models with a Euclidean
    ``external" space-time, where also $w=+1$.

    A simple analysis shows that a positive energy density $-T^t_t$ of the
    fields $F^I$ is achieved in all Lorentzian models with the signature
    $(-++\ldots +)$ if one chooses in (\ref{2.1}), as usual, $\eta_s = 1$
    for all $s$.  In more general models, with arbitrary $\eps_i$, the
    requirement $-T^t_t >0$ is fulfilled if
\bearr
     \eta_{\e I} = - \eps(I)\eps_t(I),\cm
     \eta_{\m I}= - \eps(\oI) \eps_t(\oI),       \label{2.25}\\ \lal
	\eps(I)\eqdef \prod_{i\in I}\eps_i, \cm
	\eps_t(I)= \vars{ 1, &  \R_t \subset \M_I,\\
                      -1  & \mbox {otherwise} }            \label{epst}
\ear
    where $\R_t$ is the time axis.
    If $\eps_t(I) =1$, we are dealing
    with a genuine electric or magnetic field, while otherwise the $F$-form
    behaves as an effective scalar or pseudoscalar in the physical
    subspace. The latter happens, in particular, in isotropic
    cosmologies and their Euclidean counterparts where the time coordinate
    is $u$ and $\R_t=\R_u$, unrelated to any subset $I$.
    $F$-forms with $\eps_t(I) =-1$ will be called {\sl quasiscalar.\/}

\medskip\noi
{\bf Example}: consider a \sph configuration, with $D=6$, $\M =
     \R_0\times\R_1\times S^2\times \R_4 \times \R_5$, where the coordinate
     indices 0, 1, 4, 5, refer to time, radius and two extra dimensions, 2
	and 3 to the spherical angles, respectively; thus $\R_0=\R_t$ and
	$R_1=R_u$. Then, for $\rank F=3$, the component $F_{015}$ is electric,
	$I \mapsto (0,5)$; $F_{234}$ is magnetic, $I\mapsto (0,5)$; $F_{145}$
	is electric quasiscalar, $I\mapsto (4,5)$; $F_{023}$ is magnetic
	quasiscalar, $I\mapsto (4,5)$, where the figures in parantheses are
	coordinate indices of the respective subspaces $\M_I$.  \medskip

     Let us now, as in \cite{Br73} and many later papers, choose the harmonic
     $u$ coordinate ($\nabla^M \nabla_M u = 0$), such that
\beq                                                         \label{3.1}
     \alpha (u)= \sum_{i=0}^{n} d_i \beta^i
	         \equiv d_0\beta^0 + \sigma_1(u), \cm
      			    \sigma_1(u) \eqdef \sum_{i=1}^{n} d_i\beta^i.
\eeq
     The Maxwell-like equations due to (\ref{2.1}) for the $F$-forms are
	easily integrated, giving
\bear                                                  \label{3.2}
	F_{\e I}^{uM_1\ldots M_{d(I)}}
		\eql Q_{\e I}\e^{-2\alpha - 2\olam_{\e I}\ophi}
                      \eps^{M_1...M_{d(I)}}/\sqrt{|g_I|},
		        \qquad  Q_{\e I}= \const,      \\
	F_{\m I,\, M_1\ldots M_{d(\oI)}}                  \label{3.3}
		\eql Q_{\m I} \eps_{M_1...M_{d(\oI)}} \sqrt{|g_\oI|},\qquad
		        \qquad  Q_{\m I}= \const,
\ear
     where $|g_I| = \prod_{i\in I}|g^i|$, $Q_s$ are charges
     and overbars replace summing in $a$.
     In what follows we will restrict the set $\S=\{s\}$ to such $s$ that
     the charges $Q_s\ne 0$.

	Consequently, at the r.h.s. of the Einstein equations due to (\ref{2.1})
	$R_M^N -\half \delta_M^N R = T_M^N$,
	the energy-momentum tensor (EMT) $T_M^N$ takes the form
\bearr                                                      \label{3.4}
    \e^{2\alpha}T_M^N
	  = -\frac{w}{2} \sums \ep_s Q_s^2
	      	\e^{2\sigma(I) -2\chi_s\olam_s \ophi}
		 			\diag\bigl(+1,\ \pO_I,\ \mO_\oI \bigr)
     + \frac{w}{2} \bigl(\dot\varphi{}^a)^2
	                    \diag(+1,\ \mO_{I_0}\bigr)
\ear
     where the first place on the diagonal belongs to $u$ and the symbol
     $[f]_J$ means that the quantity $f$ is repeated along the diagonal for
     all indices referring to $\M_j,\ j\in J$;
	$\sigma(I) \eqdef \sum_{i\in I} d_i\beta^i$;
	the sign factors $\ep_s$ and $\chi_s$ are
\beq
	\ep_{\e I} = -\eta_{\e I} \eps(I), \cm
	\ep_{\m I} = w \eta_{\m I} \eps(\oI);\cm      \label{3.14}
	\chi_{\e I}= +1, \cm
	\chi_{\m I}= -1,
\eeq
	so that $\chi_s$ distinguishes electric and magnetic forms.

     Let us suppose, as is usually (and reasonably)
	done in $p$-brane studies, that
     {\sl neither of $I_s$ such that $Q_s\ne 0$ contains
     the index 0}, \ that is, neither of the branes
     ``lives" in the subspace $\M_0$, interpreted as the external
	space or its subspace.
     (This means that, e.g., in the \sph case there is no electric or
	magnetic field along a coordinate sphere $\M_0 = S^{d_0}$.) Then each
	constituent EMT and hence the total EMT possess the property $T_u^u +
	T_z^z =0 $ if $z$ belongs to $M_0$.
     As a result, the corresponding combination of the Einstein equations
     has a Liouville form and is integrated:
\bear
     \ddot \alpha -\ddot \beta^0
                       \eql wK_0(d_0 -1)^2\e^{2\alpha - 2\beta^0} = 0, \nn
     \e^{\beta^0 - \alpha} \eql (d_0-1) S(wK_0,\ k,\ u),    \label{3.8}
\ear
     where $k$ is an integration constant (IC) and we have introduced the
     notation
\bear
     S(1,\ h,\ t) \eql \vars{ h^{-1} \sinh ht, \quad & h>0,\\
     			                        t,       & h=0,\\
     			     h^{-1} \sin ht,     & h<0; }    \nn
     S(-1,\ h,\ t)\eql h^{-1} \cosh ht; \cm  h> 0;            \nn
     S(0,\ h,\ t) \eql \e^{ht}, \cm\qquad    h\in \R.         \label{3.9}
\ear
     Another IC is suppressed by properly choosing the origin of the $u$
     coordinate.

     With (\ref{3.8}) the $D$-dimensional line element may be written in
     the form  ($\od \eqdef d_0-1$)
\beq
     ds^2 = \frac{\e^{-2\sigma_1/\od}}{[\od S(wK_0,k,u)]^{2/\od}}
     	  \biggl[ \frac{w\, du^2}{[\od S(wK_0,k,u)]^2} + ds_0^2\biggr]
     			+ \sumn \e^{2\beta^i}ds_i^2       \label{3.10}
\eeq

     Let us treat the remaining
	set of unknowns ${\beta^i(u),\ \varphi^a (u)}$
     as a real-valued vector function $x^A (u)$ (so that
     $\{A\} = \{1,\ldots,n\} \cup \A$) in an $(n+|\A|)$-dimensional vector
     space $V$ (target space). The field equations for $\beta^i$ and
	$\varphi^a$ can be derived from the Toda-like Lagrangian
\bearr                                                      \label{3.12}
     L=G_{AB}\dot x^A\dot x^B-V_Q (y)
     \equiv \sumn (\dot\beta^i)^2 + \frac{\dot\sigma_1^2}{d_0-1}
	          + \delta_{ab}\dot\varphi^a \dot\varphi^b - V_Q (y),\nnn
      V_Q (y) = -\sums \ep_s Q_s^2 \e^{2y_s}
\ear
     with the ``energy" constraint
\beq                                                      \label{3.16}
	E = G_{AB}\dot x^A \dot x^B + V_Q (y)
	                                =\frac{d_0}{d_0-1}K, \cm
        K = \vars {
                      k^2 \sign k, & wK_0 = 1; \\
                      k^2,         & wK_0 = 0,\ -1.       }
\eeq
     where the IC $k$ has appeared in (\ref{3.8}).
     The nondegenerate symmetric matrix
\beq                                                       \label{3.13}
       (G_{AB})=\pmatrix {
  	       d_id_j/\od + d_i \delta_{ij} &       0      \cr
	          0                         &  \delta_{ab} \cr }
\eeq
     defines a positive-definite metric in $\V$;
     the functions $y_s(u)$ are defined as scalar products:
\beq                                                       \label{3.15}
     y_s = \sigma(I_s) - \chi_s \olam_s\ophi
	   \equiv Y_{s,A}  x^A,    \cm\
     (Y_{s,A}) = \Bigl(d_i\delta_{iI_s}, \ \  -\chi_s \lambda_{sa}\Bigr),
\eeq
     where $\delta_{iI} =1$ if $i\in I$ and $\delta_{iI}=0$ otherwise.
     The contravariant components and scalar products of the vectors $\vY_s$
     are found using the matrix $G^{AB}$ inverse to $G_{AB}$:
\bearr                                                      \label{3.18}
     (G^{AB}) = \pmatrix{
	\delta^{ij}/d_i - 1/\oD &      0      \cr
	0                       &\delta^{ab}  \cr }, \cm\cm
	(Y_s{}^A) =
  \Bigl(\delta_{iI}-\frac{d(I)}{\oD}, \quad -\chi_s \lambda_{sa}\Bigr); \\
\lal  Y_{s,A}Y_{s'}{}^A \equiv \vY_s \vY_{s'}
	                  = d(I_s \cap I_{s'})                \label{3.20}
     			      - \frac{d(I_s)d(I_{s'})}{\oD}
			      + \chi_s\chi_{s'} \olam_s \olam_{s'}, \cm
     \oD = D-2.
\ear
     The equations of motion in terms of $\vY_s$ read
\beq
	\ddot{x}{}^A = \sums q_s Y_s{}^A \e^{2y_s},
	\cm    q_s \eqdef \ep_s Q_s^2.    \label{3.20a}
\eeq

\section{General properties of brane systems}  

     The positive energy requirement (\ref{2.25}) that fixes the input signs
     $\eta_s$, can be written as follows for Lorentzian models
     using the notations (\ref{3.14}):
\beq
     \ep_s = \eps_t(I_s).                      \label{4.1}
\eeq
     The corresponding requirement for Euclidean models is obtained
     by applying the conventional Wick rotation to Lorentzian cosmologies.
	This rotation of the time $t$ changes $w$ but preserves all $\eta_s$
	as well as $\eps(I)$ since $\R_t \not \subset \M_I$, $\forall I$.
	Then by (\ref{3.14}), $\ep_{\e I}$ remain invariable while $\ep_{\m I}$
	change. This distinction between electric and magnetic forms is also
	connected with the property of the duality transformation to change the
	sign of the EMT in Euclidean models \cite{GiSt,HKim}.

	Table 1 shows the sign factors $wK_0$ and $\ep_s=\sign q_s$ for
	$F$-forms in different classes of models under the above positive
	energy requirement.

\begin{center}

Table 1. \ Sign factors $wK_0$ and $\ep_s$ for different kinds of models
\medskip

\begin{tabular}{||l|l|c|c|c||}
\hline
\multicolumn{2}{||c|}{}  & Cosmology&  Static spaces &  Euclidean \mas \\
\multicolumn{2}{||c|}{}  &   $w=-1$ &    $w=+1$      &    $w=+1$  \mas \\
\hline
\multicolumn{2}{||c|}{$wK_0$}& $-K_0$ &  $K_0$       &      $K_0$ \mas \\
\hline
	    & electric    &  none    &     $+ 1$     &      none  \mas \\
\cline{2-5}
\raisebox{-1ex}[0ex][0ex]{$\ep_s$}
	    & magnetic    &  none    &      $+ 1$    &      none  \mas \\
\cline{2-5}
	    & electric quasiscalar
	                  &  $- 1$   &    $- 1$      &     $- 1$  \mas \\
\cline{2-5}
	    & magnetic quasiscalar
	                  &   $- 1$  &    $- 1$      &     $+ 1$  \mas \\
\hline
\end{tabular}
\end{center}

	In what follows, we restrict ourselves to the model described in \sect
	2 with the sign factors specified in Table 1, unless specially
	indicated.

     One general statement, to be taken into account in the subsequent
     proofs, can be formulated as a lemma:

\Theorem {Lemma 1}
     {At any regular point of the space-time,
      for all $a\in \A$ and $s\in \S$,}
\beq
	\e^{-2\alpha}(\dot{\varphi}{}^a)^2 < \infty, \cm
	\e^{-2\alpha + 2y_s} < \infty.                         \label{reg}
\eeq

\noi Indeed, regularity implies finite values of all curvature invariants,
     including $R$ and $R_M^N R^M_N$; by virtue of the Einstein equations,
     one must have $T_M^N T^M_N < \infty$.
     Since $T_M^N$ has a block-diagonal structure, the latter invariant can
     be written as a sum of squares, where each summand must thus be finite,
     including $(T_t^t)^2$. The component $T^t_t$ is in turn, due to
     (\ref{4.1}), a sum of negative-definite terms, corresponding to
     scalar fields $\varphi^a$ and $F$-forms $F_s$. Therefore every such
     term must be finite, leading to (\ref{reg}).

\subsection{Isotropic cosmology}

     Table 1 shows that in isotropic cosmologies, when $u$ is a time
	coordinate and $\M_0$ is identified with the physical space
	(conventionally $d_0=3$), $\ep_s =-1$: there are only quasiscalar forms
	since a true electric or magnetic field would violate the spatial
	isotropy.

     The logarithm of the extra-dimension volume factor,
     $\sigma_1$, by virtue of (\ref{3.20a}) obeys the equation
\beq                                                         \label{C1}
	\ddot{\sigma}_1 = -\frac{d_0-1}{D-2}\sums d(I_s)Q_s^2 \e^{2y_s},
\eeq
     whence $\ddot{\sigma}_1 <0$. So this volume factor cannot have a
     minimum and, moreover, if it tends to a finite value
     $\e^{\sigma_{10}}$ as $u\to\pm\infty$, at other values of $u$ it is
     smaller than $\e^{\sigma_{10}}$. This feature is unfavourable for
     obtaining models with the so-called dynamical compactification, where
     the size of extra dimensions decreases to microscopic scales in the
     course of the evolution.

     Next, due to $\ep_s=-1$, both terms in the expression (\ref{3.16})
     for $E$ are positive-definite, so that nontrivial solutions correspond
     to $k>0$. The range of $u$ is $\R$ for $K_0=0,+1$ and (without loss os
	generality) $u>0$ for $K_)=-1$. By (\ref{3.9}) and (\ref{3.10}),
	the model asymptotics are characterized as follows.

     For any $K_0$, at the asymptotic $u\to +\infty$ the
     total volume factor $\e^{d_0\beta^0 + \sigma_1}$ (which, by
     (\ref{3.1}), coincides with $\e^\alpha$) tends to zero. Although
     separately the physical scale factor $a(u) = \e^{\beta_0}$ and the
     ``internal" one, $\e^{\sigma_1}$, may have various limits, the
     behaviour of $\e^{\alpha}$ indicates that this asymptotic is singular.
     Moreover, since asymptotically  $\alpha \sim cu$, $c =\const <0$, the
     proper time $t = \int \e^{\alpha}du < \infty$: the singularity occurs
     at finite proper time.

     For $K_0=+1$ the other asymptotic $u\to -\infty$ is like the one
     just described, due to the symmetry of the function $\cosh ku$ in
     (\ref{3.9}). Thus closed models evolve in a finite proper time
     interval between two singularities where the total volume of the
     Universe tends to zero.

     For $K_0=0$, the asymptotic $u\to -\infty$ corresponds to an infinitely
     growing total volume factor $\e^{\alpha}$ while the proper time $t$ is
     also infinite. In the special case when $\sigma_1 \to
     \sigma_{10}=\const$, the physical scale factor $a$ obeys the law
     $a\sim |t|^{1/d_0}$.

     For $K_0=-1$ the second asymptotic is $u\to 0$, and this is a regular
     point of the equations of motion (\ref{3.20a}) determining $x^A$.
	So the metric behaviour is (now in the general case)
	governed by the function
     $S(1,k,u) \approx u$ in (\ref{3.10}), while all $\e^{\beta^i}$, $i>0$
	and consequently $\sigma_1$ tend to finite limits. As $u\to 0$,
\beq
	\e^{\alpha}\sim u^{-1-1/\od}, \qquad |t|\sim u^{-1/\od} \to \infty,
     \qquad
     	 a(t) = \e^{\beta^0} \sim |t|,                     \label{C2}
\eeq
     corresponding to linear expansion or contraction of the physical space.

     Finally, \eq (\ref{3.16}) with $V_Q>0$ implies that all
     $\dot{x}{}^A$ are bounded above, hence $x^A(u)$ are finite for all
     finite $u$ and cannot create a singularity. Therefore the above
     description of the asymptotics is quite general and applies to all
     isotropic cosmologies in the field model under consideration.

	It should be noted, however, that this discussion concerns the model
	behaviour in the $D$-dimensional Einstein conformal frame, in which the
	action (\ref{2.1}) was postulated. See further comments in \sect 6.

\subsection{Static spherical symmetry: general observations}

     In static, \sph models, where $u$ is a radial
     coordinate, $w=+1$, $\M_0=S^{d_0}$, $K_0 = +1$, among other $\M_i$
     there should be a one-dimensional subspace, say, $\M_1$, which may be
     identified with time: $\eps_1 = +1$. The sign factor $wK_0$ in
     (\ref{3.8}) is $+1$, while $\ep_s$ is, due to (\ref{4.1}), $+1$ for
     normal electric and magnetic forms $F_I$ and $-1$ for quasiscalar ones.

     By construction, see \eqs (\ref{3.9}), (\ref{3.10}), spatial infinity
     corresponds to $u=0$ (where the usual ``area function"
     $\e^{\beta^0} \sim u^{1/\od}$) and, without loss of generality, the
     range of $u$ is
\beq
	0 < u < \umx                                  \label{G1}
\eeq
     where $\umx$ is either $+\infty$, or the smallest value of $u$ where
     the fields lose regularity.

     The experience of dealing with particular models belonging to the
     class (\ref{2.1}) indicates that a generic \sph solution exhibits a
     naked singularity. Possible exceptions can be (i) \bhs\ (BHs), (ii)
     \whs\ (WHs) or \wh-like objects with a neck and a second
	nonsingular asymptotic, (iii) configurations with a regular centre (a
	soliton-like object, which might be expected for an interacting field
	system) and, finally, (iv) a situation where the coordinate patch we
	use is incomplete, terminates at a regular sphere $u=\umx$ (which may
	be even infinitely remote in our static frame of reference), and a
	possible continuation may reveal either a singularity, or one of the
	opportunities (i)--(iii).

     One can show, however, that for our model only the BH opportunity is
     viable. Lorentzian WHs do not exist according to \cite{bobs} (see also
	\sect 5), while variants (iii) and (iv) are ruled out by the following
	theorem:

\Theorem {Theorem 1}
     {The present model does not admit solutions
     describing a static, \sph configuration (a) with a regular
     centre or (b) where $u=\umx$ corresponds to a regular surface
     such that $\M_0$ is a sphere of finite radius.  }

\noi {\bf Proof. (a)} \ A regular centre implies local flatness of the
     metric at some $u = u^*$ where $\e^{\beta^0} = 0$, while other
     $\beta^i$ remain finite.  One easily shows that with (\ref{3.10}) it
     may happen only when $k=0$, $u^* = \umx = \infty$
     (otherwise the correct radius-to-circumference ratio for small circles
     around the centre cannot be achieved).	Then
     due to (\ref{3.1}), since $|\sigma_1| < \infty$,
\beq
	\e^{\beta^0} \sim u^{1/(d_0-1)}, \qquad           \label{G2}
	\e^{\alpha} \sim u^{-d_0/(d_0-1)}  \cm
	                     {\rm as}\qquad u\to\infty.
\eeq
     On the other hand, the EMT regularity requirement%
\footnote
{One might just require $|\varphi| < \infty$ as part of the centre
regularity conditions. Our proof, however, also rules out a hypothetical
situation when an infinite $\varphi$ value, due to the factor
$\e^{2\lambda\varphi}$, modifies the behaviour of $F$-forms, leading to a
regular geometry.
}
     (see Lemma 1) leads to $|\varphi| < \infty$ as $u\to\infty$. Therefore
     at such a centre the $F$-forms behave like free fields exhibiting (see
     (\ref{3.2})--(\ref{3.4})) a singularity, with infinite values of
     the EMT invariants. Item (a) is proved.

     The assumption {\bf (b)} means that both $\beta_0$ and $\sigma_1$ are
     finite at $u=\umx$. This cannot happen at $\umx<\infty$ since there
     would be no reason to stop at this value of $u$; and at $\umx = \infty$
     this means that $ S(1,k,\infty) < \infty$, contrary to the definition
     (\ref{3.9}).

\subsection{Black holes: no-hair and single-time theorems}    

     We see that the only positive-energy Lorentzian \sph configurations
     without naked singularities are BHs. BH solutions
     of various models belonging to the class (\ref{2.1}) have been studied
     in numerous recent papers (see \cite{brane,bim97,AIR} and references
     therein).  However, exact solutions have been
     (and probably can be) only obtained for a small subset of the whole
     set of models (\ref{2.1}), and it makes sense to look for general
     properties of BH solutions which may be discovered without solving the
     equations. Two such properties, having the form of restrictions
     generalizing the previously observed properties of specific solutions
     \cite{bim97,bobs}, are proved here.

     In what follows, the word ``horizon" will mean a
     nonsingular surface $u = u_*$ in $\M$ where some scale factors
     $\e^{\beta^i} =0$ (corresponding to possibly multiple time
     coordinates), while other $\beta^i$ remain finite. A BH solution is
     a static, \sph solution containing a horizon. These working
     definitions, though incomplete, are sufficient for our purposes.

     An immediate observation is

\Theorem {Lemma 2}
     {BH solutions can only exist for $k \geq 0$ and the horizon is then at
     $u=\infty$.
     }

     Indeed, at a horizon, the function $\sigma_1$ defined in (\ref{3.1})
     tends to $-\infty$ along with a part of its constituents, another
     part remaining finite. According to (\ref{3.10}), to obtain a finite
     value of $\beta^0$, one has then to require that $S(1,k,u_*)=+\infty$,
     which by (\ref{3.9}) is only possible when $k>0$ and $u_*=\infty$.

     Another result applies to BHs in manifolds $\M$ with
     several time coordinates, as suggested in some recent unification
	models (see \cite{2times,im2t} and references therein). If there is
	another time coordinate, some branes can evolve with it. The
	following theorem shows, however, that in our framework, even in a
	space-time with multiple times, a BH can only exist with its unique
	preferred, physical time, while other times are not distinguished by
	the metric behaviour from extra spatial coordinates.

\Theorem {Theorem 2 (Single-Time Theorem)}
     {Any BH solution with $k> 0$ contains precisely one coordinate $t$ such
     that $g_{tt}=0$ at the horizon.}

\noi {\bf Proof.} \
     Suppose that $u=\infty$ is a horizon where some $\e^{\beta^i} \to 0$,
     $i \in I_t \subseteq (I_0 \setminus 0)$. As follows from (\ref{3.1}),
     at the asymptotic $u\to\infty$ one has $\alpha \to -\infty$ and,
     moreover, the finiteness of $\beta^0$ means (see (\ref{3.10})) that
     $\alpha \sim -ku$. On the other hand, the condition (\ref{reg})
     holds only if for all $F$-forms, at most,
\beq
	\e^{2y_s} = O(\e^{-2ku}).                   \label{T1}
\eeq
     The equations of motion (\ref{3.20a}) then show that, as $u\to\infty$,
\beq
	\dot{x}{}^A = -c^A +o(1), \cm  c^A = \const       \label{T2}
\eeq
     where $c^i >0$ for $i\in I_t$ and $ c^i=0$ for other $A$
     (see Remark 2).

     In the constraint (\ref{3.16}), the potential
     $V_Q(u)\toas{u\to \infty} 0$ due to (\ref{T1}), therefore
\beq
	G_{AB} c^A c^B  = \frac{d_0}{d_0 -1}.    \label{T3}
\eeq
     The asymptotic of $\alpha$ and the condition (\ref{3.1}) show that,
     simultaneously,
\beq
	\sum_{i\in I_t} d_i c^i =k,                      \label{T4}
\eeq
     so that $c_i \leq k$.
     From (\ref{T3}) with (\ref{T4}) and (\ref{3.13}) it follows
\beq
	\sum_{i \in I_t} d_i {c^i}^2 = k^2.                \label{T5}
\eeq
     Combined, \eqs (\ref{T4}) and (\ref{T5}) lead to
\beq
	\sum_{i \in I_t} d_i c^i (k-c^i) =0,             \label{T6}
\eeq
     which is compatible with (\ref{T5}) for $0 \leq c^i \leq k$ only when
     the sum consists of one term, to be labelled $i=1$, such that
     $d_1=1$ and $c^1 =k$. This proves the theorem.

\medskip
     One more theorem shows that BH solutions can contain only true
     electromagnetic $F$-forms rather than quasiscalar ones.

\Theorem {Theorem 3 (No-Hair Theorem)}
     {All $F$-forms in a BH solution with $k>0$ possess the property
     $\delta_{1I_s} =1$, where the number $i=1$ refers to the time axis.}

     The proof rests on Lemma 1, which, applied to $F$-forms, leads again to
     (\ref{T1}). Now, according to Theorem 2, at a horizon ($u\to\infty$)
     only $\beta^1 \to -\infty$, while other $\beta^i$ are finite. As is
     directly verified, (\ref{T1}) holds in the case $\delta_{1I_s}=1$ (for
     true electromagnetic forms), while for quasiscalar ones one has finite
     limits for $\e^{y_s}$, leading to infiniteness of the corresponding EMT
     constituent.

\medskip\noi
{\bf Remark 1.}
     The regularity of the scalar fields, $x^A = \varphi^a$, at $u\to \infty$
     was not required in the conditions of Theorems 2 and 3;
     for $k>0$ it follows from (\ref{reg}).
     Under the additional requirement $\varphi^a < \infty$ as $u\to \infty$,
	Theorem 3 is easily proved for $k=0$ as well.

\medskip\noi
{\bf Remark 2.}
     For $k=0$ we have no Theorem 2; moreover, Theorem 3 is not proved
	for $k=0$ without assuming $\varphi^a < \infty$. Nevertheless,
     for BH solutions with $k=0$ obtainable as a limit of ones with $k>0$,
     the statements of both theorems remain valid. (For known exact BH
     solutions, $k=0$ corresponds to the extreme limit of minimal mass for
     given charges.) Meanwhile, the existence of exceptional BH solutions
	with $k=0$, nonzero quasiscalar forms and/or multi-time horizons is not
	ruled out by our study; such solutions may perhaps exist with infinite
     limits of scalar fields that balance the infinity of $\e^{-\alpha}$ in
     the EMT of $F$-forms.

\medskip\noi
{\bf Remark 3.}
     If there are BH solutions, there are also others,
     where the scale factor showing a zero value is associated,
     instead of physical time, with one of the extra coordinates
     (such solutions are obtained from BH ones by simple re-denoting). One
     thus finds the so-called T-holes, where crossing a horizon leads to
     changing the signature of the external, physical space from
     ($-+++\cdots$) to ($--++\cdots$).  Possible properties of such objects
     are discussed in more detail elsewhere \cite{br95,br-mod} within the
     frames of dilaton gravity, but the considerations thereof are valid as
     well for the more general model (\ref{2.1}).  Theorems 2 and 3 are
     valid for T-holes after proper re-formulation.

\section{Some exact solutions}  

\subsection{Orthogonal systems (OS)}

     The field equations are entirely integrated if all
     $\vY_s$ are mutually orthogonal in $\V$, that is,
\beq                                                        \label{3.21}
     \vY_s \vY_{s'} = \delta_{ss'}\big/ \Nsq, \cm
	     1\big/ \Nsq =
	d(I)\bigl[1- d(I)/\oD \bigr] + {\olam_s}^2 >0.
\eeq
     Then the functions $y_s(u)$ obey the decoupled Liouville equations
     $\ddot y_s = b_s\e^{2y_s}$, with $b_s \eqdef \ep_s Q_s^2/\Nsq$,
     whence
\beq                                                   \label{3.23}
     \e^{-y_s(u)} = \sqrt{|b_s|} S(\ep_s,\ h_s,\ u+u_s),
\eeq
     where $h_s$ and $u_s$ are ICs and the function $S(.,.,.)$ has been
     defined in (\ref{3.9}). For the sought functions
     $x^A (u) = (\beta^i,\ \varphi^a)$ we then obtain:
\bear                                                      \label{3.24}
     x^A(u) \eql \sums \Nsq Y_s{}^A y_s(u) + c^A u + \uc^A,
\ear
     where the vectors of ICs $\vec c$ and $\vec\uc$ are orthogonal
     to all $Y_s$: \ $c^A Y_{s,A} = \uc^A Y_{s,A} = 0$, or
\beq
     c^i d_i\delta_{iI_s} - c^a\chi_s\lambda_{sa}=0, \inch
     \uc^id_i\delta_{iI_s} -\uc^a\chi_s\lambda_{sa}=0.     \label{3.25}
\eeq
	The solution is general for the properly chosen input parameters;
	the number of independent charges equals the number of $F$-forms.

\subsection{Block-orthogonal systems (BOS)}

     Suppose now \cite{bobs} that the set $\S$ splits into several
	non-intersecting non-empty subsets,
\beq                                                      \label{*1}
     \S = \bigcup_{\o}\Som, \cm |\Som|=m(\o),
\eeq
     such that the vectors $\vY_{\mu(\o)}$ ($\mu(\o) \in
     S_{\o}$) form mutually orthogonal subspaces $\V_\o$ in $\V$:
\beq
     \vY_{\mu(\o)} \vY_{\nu(\o')} = 0, \cm \o \ne \o'.
		                                                  \label{*.5}
\eeq
     Suppose, further, that, for each fixed $\o$, all $\vY_\nu$
	(where $\nu=\nu(\o)$) are linearly independent and the charge
	factors $q_\nu = \ep_\nu Q_\nu^2 \ne 0$ satisfy the set of linear
	algebraic equations
\beq
     (\vY_{\nu}-\vY_{\nu'}) \vZ_\o =0,\cm
               \vZ_\o \eqdef \summo q_\mu \vY_\mu,           \label{*3}
\eeq
     for each pair $(\nu, \nu')$. Then the function
    	$y_\o(u) \eqdef Y_{\mu(\o), A}x^A$ is the same for all $\mu\in\S_\o$
	and satisfies the Liouville equation $\ddot y_\o= b_\o \e^{2y_\o}$. As
	a result, we obtain a solution to the equations of motion,
	generalizing (\ref{3.23}), (\ref{3.24}):
\bear
     \e^{-y_\o}
     \eql \sqrt{|b_\omega|} S(\sign b_\omega, h_\omega, u+u_\omega),
                                                             \label{*5}\\
                                                             \label{*6}
     x^A \eql \sumo N_\omega^2 Y_\o{}^A y_\omega(u)+  c^A u + \uc^A,
\ear
     where $h_\o$ and $u_\o$ are ICs;
	the constants $c^A$ and $\uc^A$ satisfy the same orthogonality
     relations (\ref{3.25}) as for OS, that is, the vectors $\vec c$ and
     $\vec {\uc}$ are orthogonal to each individual
     $\vY_s$, even if it is a member of a BOS subsystem.
	We have used the notations
\beq
       b_\o = \vY_{\nu(\o)}\vZ_\o;   \cm
     \vY_\o = \vZ_\o \hq\o;  \cm
  N_\o^{-2} = \vY_\o^2 = \frac{b_\o}{\hq_\o}; \cm
     \hq_\o = \summo q_\mu                                \label{*4}
\eeq
	Here $b_\o$ is nonzero and independent of $\nu(\o)\in S_\o$ due to
	(\ref{*3}); moreover, $\hq_\o \ne 0$ since $\hq_\o = \vZ_\o^2/b_\o$
	while the nonzero vector $\vZ$ is determined up to extension by
	(\ref{*3}).

	The linear independence of $\vY_{\mu(\o)}$ thus guarantees that
	\eqs(\ref{*3}) yield $q_{\mu(\o)}$ for a given subsystem up to a common
	factor. Therefore, unlike the OS solution, the BOS one is special: the
	number of independent charges coincides with $|\{\o\}|$, the number of
	subsystems; however, we thus gain exact solutions for more general sets
	of input parameters, e.g. a one-charge solution can be obtained for
	actually an arbitrary configuration of branes with linearly independent
	$\vY_\mu$ (except possible cases when the solution of (\ref{*3}) leads
	to at least one zero charge).

     When $m=1$, we have a single vector $\vY_\o =\vY_s$ orthogonal
	to all others, with the norm $N_\o^{-2} = N_s^{-2}$, and the charge
	factor is $b_\o=b_s$. Thus single branes and BOS subsystems are
	represented in a unified way, and the OS solution is a special case
	($m(\o)=1,\forall \o$) of the BOS one.

     The metric has the form (\ref{3.10}), where the function $\sigma_1$ is
\beq                                                       \label{*8}
     \sigma_1 = - \frac{d_0-1}{D-2}
     		\sumo N^2_{\o}y_\o(u)
     \summo \frac{q_\mu}{\hq_\o}d(I_\mu) + u\sumn c^i + \sumn \uc^i.
\eeq
     For OS ($\o \mapsto s$) the sum in $\mu$ reduces to $d(I_s)$.
	The ``conserved energy" (\ref{3.16}) is
\beq
     E = \sumo N_\omega^2                                  \label{*7}
	       h_\omega^2\sign h_\omega +  c_A c^A
	                                = \frac{d_0}{d_0-1}.
\eeq

     In the special case $m=2$, $\vY_1^2 = \vY_2^2$, one easily obtains
     $b_1=b_2$, as was shown in \cite{Pd97} for a single $F$-form. By
     definition of $b_s$ that means not only $Q_1^2=Q_2^2$, but also a
     coincidence of the sign factors $\sign b_s = \ep_s$. For
     instance, in spherical symmetry, the $F$-fields must be either both
     true electric/magnetic ones ($\sign b_s=1$), or both quasiscalar ones
     ($\sign b_s= -1$).

\subsection{On cosmological and black-hole solutions}

	There is a large number of exact cosmological solutions to
	special cases of the model (\ref{2.1}), see \cite{Greb,LuMaha,KKO} and
	references therein. It can be seen that the description
	of \sect {3.1} (which is certainly confirmed by exact solutions)
	actually exhausts all general features of the model, since other
	details, such as, e.g., the particular behaviour of the physical scale
	factor $a(t)$, depend on the choice of integration constants.

     BHs are obtained as special \sph solutions when $h_\o >0$,
     $\umx=\infty$. The functions
     $\beta^i$ ($i=0,2,\ldots,n$) and $\varphi^a$ remain finite as
     $u\to\infty$ under the following constraints on the ICs:
\bear
     h_\o = k, \cm \forall\  \o; \inch
     c^A = k \sumo \Noq Y_\o{}^A - k \delta^A_1            \label{5.4}
\ear
     where $A=1$ corresponds to $i=1$ (time), $d_1=1$ (according to Theorem
     2). The constraint (\ref{*7}) then holds automatically.

     The subfamily (\ref{5.4}) exhausts all BH solutions under OS or BOS
     assumptions, except the extreme case $k=0$; extreme BHs are obtained
	by subsequently passing to the limit $k\to 0$. One can notice that
	exceptional extreme BH solutions, whose possibility was mentioned in
	\sect {3.3}, are not found in this way.

     General explicit forms of OS and BOS BH solutions have been presented in
     Refs.\,\cite{bim97} and \cite{bobs}, respectively. The BH properties
     stated in Theorems 2 and 3 are confirmed for the OS and BOS
	solutions and, moreover, have been first observed
	\cite{bim97,bobs} for these solutions.

\section {Wormholes}

\subsection {Wormhole existence conditions}

	Wormhole-like configurations which can appear as special OS or BOS
	solutions, have an infinite ``external
	radius" $\e^{\beta^0(u)}$ at both ends $u_{\pm}$ of the $u$ range
	are regular between them; all $\beta^i (u_{\pm})$ ($i>0$) and
	$\varphi^a (u_\pm)$ are finite. This happens when $k<0$ and the
	solution behaviour is governed by the function $\sin ku$ (so that
	$u_-=0$ and $u_+ = \pi/|k|$) and is possible if the first positive zero
	of the function $\sin [|h_s|(u-u_s)]$ is greater than $\pi/|k|$ for any
	$s$ such that $h_s<0$ --- see Fig.\,1.

	In the cosmological setting, this behaviour would correspond to
	nonsingular, bouncing models, which are, however, absent according to
	\sect {3.1} (due to $k>0$). The static and Euclidean cases are not
	{\it a priori\/} excluded.

     As is evident from Fig.\,1, any WH solution is characterized by
     $|k| > |h_\o|$ for all $h_\o$ which are negative.
     Due to (\ref{*7}), for $k<0$ at least some $h_\o$ should be negative
     as well. Furthermore, for $k<0$ and $h_\o<0$ it is necessary to have
     $wK_0=1$ and $b_\o > 0$, respectively.

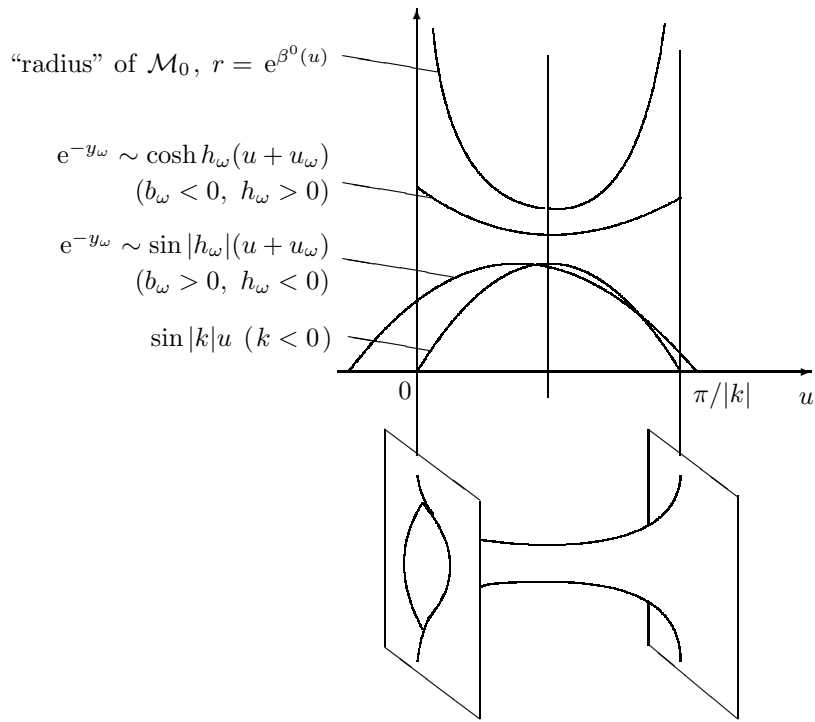
\begin{figure}
\centering


\unitlength=0.7mm
\linethickness{0.4pt}
\begin{picture}(145.00,149.00)
\put(55.00,80.00){\vector(1,0){90.00}}
\bezier{384}(70.00,80.00)(95.00,121.00)(120.00,80.00)
\bezier{420}(57.00,80.00)(88.00,121.00)(123.00,80.00)
\bezier{236}(70.00,115.00)(93.00,98.00)(120.00,113.00)
\bezier{200}(73.00,145.00)(76.00,113.00)(94.00,111.00)
\bezier{216}(95.00,111.00)(112.00,109.00)(117.00,146.00)
\bezier{152}(94.00,47.00)(119.00,47.00)(120.00,60.00)
\bezier{160}(120.00,25.00)(120.00,40.00)(95.00,40.00)
\put(114.00,51.00){\line(0,1){18.00}}
\put(114.00,69.00){\line(4,-3){17.00}}
\put(131.00,56.33){\line(0,-1){42.33}}
\put(131.00,14.00){\line(-6,5){17.00}}
\put(114.00,28.00){\line(0,1){8.00}}
\put(64.00,69.00){\line(4,-3){18.00}}
\put(82.00,55.33){\line(0,-1){41.33}}
\put(82.00,14.00){\line(-4,3){18.00}}
\put(64.00,27.67){\line(0,1){41.33}}
\bezier{116}(71.00,55.00)(81.00,43.00)(72.00,33.00)
\put(144.00,76.00){\makebox(0,0)[ct]{$u$}}
\put(54.00,86.00){\makebox(0,0)[rc]{$\sin |k|u$ ($k<0$)}}
\put(73.00,84.00){\line(-6,1){16.00}}
\put(77.00,98.00){\line(-6,1){21.00}}
\put(73.00,113.00){\line(-6,1){17.00}}
\put(74.00,137.00){\line(-6,1){18.00}}
\put(54.00,104.00){\makebox(0,0)[rc]{$\e^{-y_\o}\sim\sin |h_\o|(u+u_\o)$}}
\put(54.00,97.00){\makebox(0,0)[rc]{$(b_\o > 0,\ h_\o<0)$}}
\put(54.00,121.00){\makebox(0,0)[rc]{$\e^{-y_\o}\sim\cosh h_\o(u+u_\o)$}}
\put(54.00,114.00){\makebox(0,0)[rc]{$(b_\o < 0,\ h_\o >0)$}}
\put(54.00,139.00){\makebox(0,0)[rc]{``radius" of $\M_0$,                                      
            $r = \e^{\beta^0 (u)}$}}
\put(122.00,78.00){\makebox(0,0)[lt]{$\pi/|k|$}}
\put(69.00,78.00){\makebox(0,0)[rt]{0}}
\put(95.00,75.00){\line(0,1){65.00}}
\put(120.00,75.00){\line(0,1){66.00}}
\bezier{52}(95.00,47.00)(89.00,47.00)(82.00,48.00)
\bezier{32}(70.00,60.00)(71.00,55.00)(73.00,53.00)
\bezier{40}(70.00,25.00)(71.00,32.00)(73.00,34.00)
\bezier{52}(95.00,40.00)(83.00,40.00)(82.00,39.00)
\bezier{112}(71.00,55.00)(64.00,44.00)(71.00,31.00)
\put(70.00,64.00){\vector(0,1){85.00}}
\put(120.00,75.00){\line(0,-1){11.00}}
\end{picture}

 \caption{A wormhole configuration: qualitative picture}
\end{figure}

     Table 1 shows that WHs can exist in static or Euclidean models only
     with spherical rather than pseudospherical or planar symmetry. In
     cosmology we have no fields capable to give negative $h_s$ or $h_\o$,
     which again confirms the absence of nonsingular ``bounced" models.
     In static spherical symmetry the necessary $F$-forms are true
     electric and magnetic ones. In Euclidean models, magnetic quasiscalar
	forms are needed.

     Suppose $k<0$. Since in $\vec c\,{}^2\geq 0$,
     the requirement $|k| > |h_s|$ means that
\beq                                                   \label{4.2a}
     \sum_{\{\o:\ h_\o <0\}}  \Noq > \frac{d_0}{d_0-1}
\eeq

     This inequality is not only {\sl necessary\/}, but also
     {\sl sufficient\/} for the existence of WHs with given input
     parameters: $d_i$ and the vectors $\vY_s$. Indeed,
     put $c^A=0$ and turn to zero the charges $Q_{\mu(\o)}$ in all
	subsystems with $\hq_\o<0$ (note that, by (\ref{*4}), $\sign \hq_\o =
	\sign b_\o$.) Choose all $h_\o$ to be negative and equal, then due to
	(\ref{4.2a}) $|h_\o| < |k|$. It is now an easy matter to choose the ICs
	$u_\o$ in such a way that $\sin [|h_\o|(u+u_\o)] > 0$ on the whole
	segment $[0,\pi/|k|]$ --- and this results in a WH solution.

\subsection
{Lorentzian wormholes and a universal restriction for brane systems}

	In general relativity static \cite{HV} and even dynamic \cite{Hay}
	traversable WHs are known to violate the null energy condition.
	It can be verified \cite{bobs} that, under the present positive energy
	requirement, the model (\ref{2.1}) after reduction to $d_0+2$
	dimensions by integrating out all $\M_i$, $i>1$
	and a transition to the Einstein conformal frame, reduces to
	general relativity with a set of material fields whose EMT
	satisfies the null energy condition, which rules out static WHs.
	On the other hand, given a static WH in $D$ dimensions as described in
	the previous subsection, it would also appear as a static WH in
	$d_0+2$-dimensional Einstein frame since the relevant conformal factor
	(the volume factor of extra dimensions) is everywhere finite and
	nonzero. We have to conclude that static WHs are absent in our model.

	This means in turn that the sufficient condition
	(\ref{4.2a}) must be violated, and a properly formulated opposite
	inequality must hold. We arrive at the following theorem
	for brane systems having an orthogonal subsystem:

\Theorem {Theorem 4}
    {Consider a vector space $\V$, with a scalar
	product defined by the metric
     (\ref{3.13}), where $d_i \in \N$, $i=0,\ldots,n$,
     $d_0>1$, $d_1=1$, $\oD=\sum_{i=0}^{n}d_i-1$, and a set of nonzero
     vectors $\vY_s$, $s \in \S$, defined in (\ref{3.15})
     ($I_s \subseteq \{1,\ldots,n\}$, $\chi_s\lambda_{sa}\in\R$).
	Let there be a subset $\S_{\bot} \subset \S$ such that
	$\vY_s \vY_{s'} =0$ for $s\ne s'$, $s,s' \in \S_{\bot}$.
	Then the following inequality holds:
\bearr
     \sum_{s\in \S_{\bot}}
	          \delta_{1I_s}\Nsq \leq \frac{d_0}{d_0-1},    \label{4.7a}
	\cm	\mbox{or for $\lambda_{sa}=0$:}                   \nnn
     \sum_{s\in \S_{\bot}}
	          \delta_{1I_s} \biggl[ d(I_s)
	                  \biggl(1-\frac {d(I_s)}{D-2}\biggr)\biggr]^{-1}
	                                  \leq \frac{d_0}{d_0-1}.
\ear
	}

	The factor $\delta_{1I_s}$ in (\ref{4.7a}) excludes quasiscalars.
	For $\S_{\bot}=\S$ the theorem has been already proved by the above
	reasoning. If there are $\vY_s \not\in \S_{\bot}$, their influence can
	be ruled out by turning to zero the corresponding charges $Q_s$, and
	then, as before, assuming the contrary of (\ref{4.7a}), we immediately
	obtain a Lorentzian WH solution.

\medskip\noi
	{\bf Comment.}
     The formulation of Theorem 4 does not mention $F$-forms, time, or any
	other physical entities and is actually of purely geometric (or
	even combinatorial) nature.  From the combinatorial viewpoint it is
	essential that in the set $I_0= \{0,\ldots,n\}$ there is a
	distinguished number, in our case 1, with $d_1=1$, included in all
	subsets $I_s$ entering into the sum.
	Our proof, however, rests on physically motivated analytical
	considerations.

\medskip
	A similar theorem for a brane system with a BOS subsystem is readily
	obtained:

\Theorem {Theorem 4a}
     {Consider the model described in \sect 2,
	under the conditions specified in the first sentence of Theorem 4.
	Let there be a subset $\S' \subset \S$ such that
	the vectors $\vY_s$, $s \in \S'$
     form a block-orthogonal system with respect to
     the metric (\ref{3.13}). Then the following inequality holds
	for $s\in \S'$:
\beq                                                   \label{4.11}
     \sum_{\{\o:\ \hq_\o > 0\}}  \Noq \leq \frac{d_0}{d_0-1}
\eeq
     where $\hq_\o$ and $\Noq$ are defined in (\ref{*4}) and, for all
     $q_s$ included in the sum,  $\ep_s = \sign q_s = -1 + 2\delta_{1I_s}$.
	}

	According to the latter condition, $\ep_s$ depend on the inclusion or
	non-inclusion of the distinguished one-dimensional factor space $\M_1$
	($= \R_t$ in Lorentzian models) into the world volume of specific BOS
	members. Thus, unlike the OS case, the sum may include $F$-forms with
	different $\ep_s$, but in such a way that the combined factor $\hq_o =
	\sum_{\mu\in\o}q_\mu$ be positive for each $\o$.

\section{Concluding remarks} 

{\bf 1.}
	Some general restrictions on the behaviour of brane
	systems described by the action (\ref{2.1}) have been obtained,
	independent of specific space-time symmetry and signature:
	cosmological asymptotics, some BH properties and a universal
	restriction on the parameters of possible othogonal or block-orthogonal
	subsystems (Theorems 4 and 4a).

	Throughout the paper, the $D$-dimensional Einstein (D-E) conformal
	frame was used, although in such a general setting of the problem
	there is no evident reason to prefer one frame or another.  For
	any specific underlying theory that leads to (\ref{2.1}) in a weak
	field limit, two conformal frames are physically distinguished: one
	where the theory is originally formulated and another, providing the
	weak equivalence principle (or geodesic motion) for ordinary matter in
	4 dimensions; the latter depends on how fermions are introduced in the
	underlying theory \cite{br95,Dick,Stmel}. The first one
	should be used when discussing such issues as singularities or topology
	of a model, etc.  (what happens as a matter of fact), while the second
	one, the so-called atomic system of measurements, is necessary for
	formulating observational predictions (what seems to us). They are,
	generally speaking, different.

	Among the present results, however, only cosmological ones are
	conformal frame-dependent if different frames are connected by
	exponentials of the internal scale factors $\beta^i$ and dilatonic
	fields $\varphi^a$. Indeed, such factors, being regular everywhere
	including horizons and asymptotics, cannot change the BH or WH nature
	of a given metric. (The only exceptions are hypothetic exceptional
	extreme BH solutions mentioned in Remark 2.)

     The conclusions of \sect {3.1} on cosmological asymptotics are directly
	applicable to theories formulated in the outset in the D-E frame, like
	the weak-field bosonic sector of $D=11$ supergravity following from
	M-theory \cite{brane}, where the action (truncated by neglecting the
	Chern-Simons term) has the form (\ref{2.1}) with a single antisymmetric
	4-form and no scalar fields.

\medskip\noi
{\bf 2.}
	Unlike Lorentzian ones, Euclidean WHs (EWHs) are not ruled out, and the
	reason is (taking, say, OS solutions as an example) that, when
	selecting the $F$-forms (branes) for WH construction, in the Euclidean
	case we are no more restricted to $I_s$ containing a distinguished
	number, connected with $\R_t$, while now $\R_t = \R_u$.  So there is a
	wider choice of $I_s$ able to give $h_s<0$ and to fulfil the WH
	necessary and sufficient condition (\ref{4.2a}).

     As seen from Table 1, EWHs corresponding to
     (\ref{2.1}), if any, may be built only with the aid of
     magnetic forms $F_s$, though the existence of electric forms in a WH
     solution is not excluded.

	The situation is well exemplified for $D=11$ supergravity.
	Indeed, the orthogonality conditions (\ref{3.21})
     are satisfied by 2-branes, $d(I_s)=3$, and 5-branes, $d(I_s)=6$,
     if the intersection rules hold:
\beq
     d(3\cap 3) =1, \cm d(3\cap 6)=2, \cm d(6\cap 6)=4.      \label{ex01}
\eeq
	(the notations are evident); for all $F$-forms $\Nsq =1/2$.
	In particular, with $d_0=2$ or $d_0=3$ and other $d_i=1$,
     there is a maximal OS of seven 2-branes \cite{Greb,bim97}:
\beq
		 \begin{array}{ll} \dst
		 a: & 123, \\
		 b: & 147,    \\
		 c: & 156,  \end{array}  \cm
		 		\begin{array}{ll} \dst
	    		 			d: & 345, \\
			               e: & 246, \\
						f: & 257, \\
						g: & 367,
		                 \end{array}                      \label{ex02}
\eeq
	where the figures $1,\ldots,7$ label 1-dimensional factor spaces,
	and for static models ``1'' refers to the time axis $\R_t$.
	Only three of these $I_s$ ($a,b,c$)
	have $\delta_{1I_s}=1$, i.e., describe true electric
     or magnetic fields in a static space-time.
     Lorentzian WHs are absent since (\ref{4.2a}) requires $\sums\Nsq >2$
     for $d_0=2$ and $>3/2$ for $d_0=3$.

	In the Euclidean case we can have
     as many as 7 magnetic 2-branes, each with $\Nsq=1/2$, and WHs are
     easily found. Though, the latter is true if one considers
	$\Fmi$ of rank 7. If one remains restricted, as usual, to
	$F_s$ of rank 4 \cite{brane}, then for magnetic forms $d(I)=6$,
	and EWHs cannot be obtained. Examples of EWHs have
	been found \cite{bobs} for $D=12$ theory \cite{khv}.

     By construction, classical EWHs possess finite actions
     and are related to possible quantum tunneling processes.
     Explicit expressions for their action and throat radii in the case of
     symmetric WHs described by OS and BOS solutions, have been calculated
	\cite{bobs} explicitly in a general form for WHs which are symmetric
	with respect to their throats.

\medskip\noi
{\bf 3.}
	The present conclusions rest on the positive energy requirement that
	seems quite natural as long as we deal with classical fields.
	Thus, in particular, the well-known singularity theorems
	of general relativity actually work as well in multidimensional
	$p$-brane cosmology. Meanwhile, the low energy limit of the
	unification theories is believed to work at scales from Planckian to
	subatomic and in the early univerrse where
	quantum effects of both gravity and material fields must be of
	importance (e.g. the Casimir effect due to compactification of extra
	dimensions), and a classical treatment is only a tentative, though
	necessary, stage in studying such systems. One can mention some papers
	discussing the relevant quantum effects: the Wheeler-DeWitt equation
	for $p$-branes \cite{IM6,LuMaha}
	and the Casimir effect in cosmology \cite{GKZhuk}.
	Some non-quantum effects able to prevent a cosmological singularity
	are discussed by Kaloper et al. \cite{KKO} and Gasperini \cite{Gasp98};
	see also references therein. All such effects necessarily violate the
	usual energy requirements and can therefore create traversable
	Lorentzian \whs.

\Acknow
 {I am grateful to V. Ivashchuk, V. Melnikov and J. Fabris for helpful
 discussions, and to colleagues from DF-UFES, Vit\'oria, Brazil,
 where part of the work was done, for warm hospitality.  The work was
 supported in part by CAPES (Brazil), by the Russian State Committee for
 Science and Technology, and by Russian Basic Research Foundation.}

\end{document}